\documentstyle[aps,eqsecnum,epsf,preprint]{revtex}
\input epsf.tex
\def\.{\cdot}
\def\e{\epsilon}
\def\o{\omega}
\def\be{\begin{eqnarray}}
\def\s{\lambda}
\def\ee{\end{eqnarray}}
\def\h{{1\over 2}}
\begin{document}
\title{Bose-Einstein Interference 
and Factorization at High Energies\cite{wu}}         
\author{C.S. Lam\cite{email}}
\address{Department of Physics, McGill University,
 3600 University St., Montreal, QC, Canada H3A 2T8}        
%\date{\today}        
\maketitle
\begin{abstract}
We consider the emission or absorption 
of $n$ identical bosons from an energetic
or a massive particle, in which the bosons and 
the source particle are allowed to be offshell.
The Bose-Einstein symmetrized amplitude can be
decomposed into sums and products of more elementary
objects which we choose to call `atoms'. The origin
of this decomposition, its significance and some of
its applications will be summarized.
\end{abstract}
 
\section{Introduction}  
It is a great pleasure for me to dedicate this
article to Prof.~Ta-You Wu. I first met Prof.~Wu many
years ago, when I was an 
undergraduate student and he was the head of the Theory Group
in the National Research Council of Canada. 
His important contributions to physics, to the training
of physicists, and his seminal role in the development of science and physics in China, 
can be found elsewhere so they will not be repeated here.
On this happy occasion of his 90th birthday, 
the Physical Societies of Chinese Mainland, Taiwan,   
Hong Kong, and Overseas have jointly organized a Meeting
in his honour. This Meeting was held in Taipei from
August 11 to 15, 1997.
I was fortunate to be involved in the
organization of this five-day meeting, and am happy
to report that over three hundred physicists from 
all over 
the world participated, including four ethnic Chinese
Nobel Laureates and a Field Medalist, as well as
many leading physicists,  especially those of Chinese ancestry.
Twelve plenary talks and more than two hundred
papers in the parallel sessions were delivered.
The success of this Meeting
can undoubtedly be attributed to the great admiration
and respect the participants have for Prof.~Wu, which
motivated them to come from afar to dedicate their work
to him.

I will base this article on the talk I gave at that
conference. It is about the `magic' of Bose-Einstein (BE)
symmetry in the context of high-energy scattering. 
The `magic' stems from a {\it factorization} (or {\it decomposition})
theorem which I will discussed below.

The experimental
demonstration of BE condensation in the last
two years \cite{BE} revives the interest in this old but important
subject. 
The condensation can be traced to an effective 
attraction of bosons at short distances,
induced by the
 {\it constructive interference}
in the symmetrization of
bosonic wave functions. These effects are well known and
I will not dwell on them any further. Instead, I would
like to ask whether BE {\it destructive interference}
plays any role in physics, 
and if so where does it become important.
Certainly not at high temperatures, where phase informations
are lost, but interestingly it 
manifests itself in high-energy scattering processes.
In fact, the presence of such destructive
interference is crucial in preventing certain theoretical
disasters to occur. The more bosons there are the more important
this interference effect will be. 
I shall come back 
to illustrate what I mean by this with some examples.

Constructive and destructive interferences
are often two sides of the same coin. Consider 
for example a simple system
of two particles with
product wave function
$\phi_1(\vec x_1)\phi_2(\vec x_2)$, where $\phi_1$ and
$\phi_2$ are normalized and orthogonal to each other. 
When BE symmetrized,
the normalized wave function becomes
$\Phi(\vec x_1,\vec x_2)=[\phi_1(\vec x_1)\phi_2(\vec x_2)+
\phi_1(\vec x_2)\phi_2(\vec x_1)]/\sqrt{2}$. At $\vec x_1
=\vec x_2=\vec x$, 
$\Phi(\vec x,\vec x)=\sqrt{2}\phi_1(\vec x)\phi_2(\vec x)
>\phi_1(\vec x)\phi_2(\vec x)$,
producing a {\it constructive interference}, 
which in the case
of many-body wave functions leads to the effective
attraction and condensation observed at low temperatures.
Since both $\Phi$ and $\phi_1\phi_2$ are normalized,
to preserve probability
this enhancement at $\vec x_1=\vec x_2$ 
must produce a depletion somewhere else, and this 
is just the
{\it destructive interference} mentioned earlier as
being important in high-energy scatterings.

It should be clarified at this point that
`high energy' means high {\it total} energy, and not 
necessarily high kinetic
energy. Large mass in the presence of low
kinetic energy would qualify as high energy as well. The latter occurs
in systems involving heavy baryons and heavy quarks. The second example
discussed below is of this latter variety.

Let me now discuss two examples illustrating the importance of BE
destructive interference at high energies.

Consider first  
high-energy ($\sqrt{s}$) near-forward
elastic scattering. Since each loop-integration can potentialy produce
a $\ln(s)$ factor, 
the effective coupling constant is
$g_{eff}^2=g^2\ln(s)$, which can be large even when the 
coupling $g^2$ is small. Consequently at high energies
perturbation diagrams of all orders are to be included, resulting in many
gauge bosons being exchanged. BE symmetrization of these virtual
bosons is carried out by summing Feynman diagrams in which
gauge vertices are permuted in all possible ways.
The theoretical effect of BE symmetry can therefore be seen by 
comparing the sum with the individual Feynman diagrams.
As mentioned before,
individual diagrams can grow with a positive power
of $\ln(s)$, and the power generally increases with the order 
of the perturbation diagram. This can easily lead to
the violation of Froissart bound!
Fortunately, in electron-electron scattering
via multiple-photon exchanges, such a
disaster is averted, and unitarity restored, when
the Feynman diagrams are added together. This is so because {\it all}
positive powers of $\ln(s)$ are cancelled in the sum. 
{\it What causes such a 
magical cancellation?} As we shall discuss later, it is the BE
destructive interference.

In QCD things become considerably more complicated. Explicit calculation
up to the sixth order \cite{CW} 
shows that, depending on the colour of the exchange channel,
some but not all 
of these powers of $\ln(s)$ may be cancelled. In fact, it turns out
that none of the $\ln(s)$ powers are cancelled in the colour-octet
channel. Instead, their contributions
pile up to form a reggeized gluon \cite{CW,REG}, thereby using another
mechanism to restore unitarity.
This is in sharp contrast to the case of QED 
discussed above, 
where photons do not reggeize
and the restoration of unitarity relies on cancellations. 

{\it Why do we have this drastic difference between two gauge theories?}
The answer can be found in 
the different way BE symmetry is being implemented in the two cases. 
Gluons carry a new
{\it nonabelian} quantum number, `colour', so for them  both
 colour and spacetime coordinates
must be BE symmetrized. This produces the qualitative difference
between QCD and QED noticed above, but it also
makes nonabelian BE symmetrization considerably
more difficult to analyse \cite{LL1,FHL,FL1,L1}.

I will now discuss briefly a second example which deals with {\it real} 
rather than {\it virtual} bosons.
In the pi-nucleon Yukawa theory derived from QCD
with a large number of colours $N_c$, the Yukawa coupling constant
is proportional to $\sqrt{N_c}$, so every
Feynman tree diagram with $n$ pions grows like $N_c^{n/2}$ \cite{NC}. 
One might conclude from this that the theory favours the production
of a large number of pions, and that high-order loop diagrams are terribly
important.
Fortunately this is not so because in summing the $n!$
tree diagrams to enforce BE symmetry,
the large amplitudes coming
from individual diagrams get quite thoroughly cancelled out, 
leaving behind only a residue
going down with $N_c$ like $N_c^{1-n/2}$ \cite{NC}. 
This cancellation once again
can be attributed to {\it BE destructive interference}
\cite{LL2}. Incidentally, this is a `high-energy' process even 
though the pions
carry little kinetic energy because the mass of the nucleon is
proportional to $N_c$.

These two examples illustrate the {\it theoretical}
 consequences of BE
destructive interference. 
I believe these interferences also have directly observable
{\it phenomenological} consequences, but unfortunately I have not 
had time to work them out as yet.

The destructive interference phenomenon discussed above follows from 
a far more fundamental property: the high-energy BE-symmetrized
tree amplitude
satisfies a {\it factorization ({\rm or} decomposition) theorem} 
\cite{LL1,FHL}.
This theorem allows the amplitude to be
decomposed into sums of products of more primitive objects 
which I choose to call
{\it atoms}. An atom may have any number of bosons, and it differs
from a Feynman diagram only in that
it carries the {\it `adjoint colour'}. Discussions on its precise 
meaning, as well as the significance 
of this decomposition, will be postponed to later sections.

For abelian amplitudes this theorem simply degenerates into the well
known {\it `eikonal formula'} \cite{EIK}. We may therefore think
of this decomposition as a {\it nonabelian eikonal formula}.

In the following sections we shall disucss these and other topics in a more 
quantitative way.

\section{High-Energy Approximation}

Consider the tree amplitude of Fig.~1 in which all 
components of the bosonic momenta $k_i$ are much less than the energy
$p^0$ of the source particle. In what follows the final momentum $p$ is 
always taken to be onshell, $p^2=m^2$, but the initial momentum
$p'=p+\sum_{i=1}^nk_i$ may or may not be. The bosonic momenta
are always arbitrary so they can be
sewed up with other bosons to turn the tree diagram into
a loop diagram. In that way the formulas developed below for tree
amplitudes can be applied to loop diagrams as well.

In this
kinematical regime, we can approximate the denominator
of a propagator by ignoring the quadratic term  $K^2$
in the expression $(p+K)^2-m^2+i\e\simeq 2p\.K+i\e\equiv
\o+i\epsilon$, where $K$ is a sum over a number of $k_i$'s.
We shall refer to this approximation as the
{\it eikonal approximation}. It can be shown that
this amounts to ignoring the
effect of recoil and treating the trajectory of the source
particle in configuration space
as a straight line throughout. This approximation
will be used in the rest of the
discussions, with $\o_i\equiv 2p\.k_i$.

We shall also assume the absence of numerators for the propagators
of Fig.~1, so if $\s_i$ represents
the $i$th vertex, the offshell  
scattering amplitude for
Fig.~1 is given by
\be
A^*[123\cdots n]&=&\prod_{i=1}^n{1\over 
\sum_{j=1}^i\o_j+i\e}\.\s_1\s_2\s_3\cdots \s_n\nonumber\\
&\equiv& a^*[123\dots n]\.\s[123\cdots n].
\label{offshell}\ee

For easy reference, we shall generically refer to the vertex factors
$\s_i$ as `colour matrices' and the nonabelian quantum numbers carried
by the bosons as `colours'. There is no implication in this terminology
that $\lambda_i$ has to be an $SU(3)$ matrix, nor even that they have
to be the generators of any group.

For the onshell amplitude, $(p')^2=m^2$ leads to
$\sum_{i=1}^n\o_i=0$. The last propagator is absent
but it is convenient to include explicitly in the
amplitude the onshell
$\delta$-function, thus making
\be
A[123\cdots n]&=&-2\pi i\delta\left(
\sum_{i=1}^n\o_i\right)\prod_{i=1}^{n-1}{1\over 
\sum_{j=1}^i\o_j+i\e}\.\s_1\s_2\s_3\cdots \s_n\nonumber\\
&\equiv& a[123\dots n]\.\s[123\cdots n].\label{onshell}\ee

\begin{figure}
\vskip -23cm
\centerline{\epsfxsize 4.7 truein \epsfbox {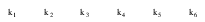}}
\nobreak
\vskip +3cm\nobreak
\caption{ (a) A Feynman
tree diagram for the emission of $n$ bosons from an energetic 
or massive particle
of final momentum $p$ and initial momentum $p'$}
\end{figure}

If $\{s\}=\{s_1s_2s_3\cdots s_n\}$
is a permutation of $\{123\cdots n\}$, 
$[s]$ is a Feynman diagram with the $n$ boson lines
similarly permuted, then its offshell amplitudes
$A^*[s]$ and its onshell amplitude $A[s]$ can
be obtained from (\ref{offshell}) and (\ref{onshell})
respectively by making this same permutation. BE 
symmetrization of the bosons are carried out by summing
the $n!$ permuted  Feynman diagrams over the symmetric group $S_n$: 
\be
M^*_n&=&\sum_{\{s\}\in S_n}^{n!}A^*[s],\nonumber\\
M_n&=&\sum_{\{s\}\in S_n}^{n!}A[s].\label{sum}
\ee

The absence of a numerator factor in the propagator
does not
mean that the source particle cannot carry a spin.
To illustrate this point let us assume the source
particle to carry spin ${1\over 2}$, and a momentum
$p+K$. In the high energy approximation,
the numerator factor $m+\gamma\.(p+K)\simeq m+\gamma\.p$
is equal to $\sum_\lambda u_\lambda(p)\bar u_\lambda(p)$, so
if $\Gamma_i$ is the true vertex factor for the diagram,
it is equivalent to a theory without the numerator factor,
but with $(\s_i)_{ab}=\bar u_a(p)
\Gamma_i u_b(p)$ as the vertex.
For example, for vector coupling ($\gamma^\mu$) to a spin-1 particle,
the effective vertex $\s$ is proportional to $\bar u_a(p)
\gamma^\mu u_b(p)=2p^\mu\delta_{ab}$. 
Physically this simply says that
the spin current is negligible compared to the translational
current in the high energy limit. Similarly, for axial vector
coupling ($\gamma^\mu\gamma_5\vec\tau\partial_\mu$) to a pion, 
the effective vertex $\s$ is proportional to $\sigma^i\vec\tau k_i$,
so in this case `colour' is really spin and isospin.
 Similar arguments can be applied
 to source particles of higher spins. In this way we are
again invoking the familiar statement that spin is 
in some sense immaterial at high
energies.

\section{Abelian Bosons}
For abelian bosons we may assume all $\s_i=1$, so $A[s]=a[s]$
and $A^*[s]=a^*[s]$. Let us first look at the simplest case,
with two bosons.
\subsection{$n=2$} 
In this case
\be
M^*_2&=&a^*[12]+a^*[21]={1\over\o_1+i\e}{1\over\o_1+\o_2+i\e}
+{1\over\o_2+i\e}{1\over\o_1+\o_2+i\e}\nonumber\\
&=&{1\over\o_1+i\e}{1\over\o_2+i\e},\label{off2}\\
M_2&=&a[12]+a[21]=-2\pi i\delta(\o_1+\o_2)\left\{
{1\over\o_1+i\e}+{1\over\o_2+i\e}\right\}\nonumber\\
&=&(-2\pi i)^2\delta(\o_1)\delta(\o_2).\label{on2}
\ee
Note that {\it factorization} occurs in both the offshell and the onshell
amplitudes. Moreover, for onshell amplitudes, the result of adding
two $1/\o$ distributions is to produce a sharply peaked {\it interference
pattern} $\delta(\o)$, with constructive interference at
$\o=0$ and complete destructive interference at $\o\not=0$.
The more $\delta$-functions there are, the more sharply peaked the
interference pattern will be. For that reason we shall measure
the `amount of spacetime interference' by the number of $\delta$-functions
present.

\subsection{Arbitrary $n$}
{\it Factorization} persists to arbitrary $n$. The result is \cite{EIK}
\be
M_n^*&=&\sum_{\{s\}\in S_n}^{n!}a^*[s]=\prod_{i=1}^n{1\over\o_i+i\e},\label{offn}\\
M_n&=&\sum_{\{s\}\in S_n}^{n!}a[s]=\prod_{i=1}^n
\left[-2\pi i\delta(\o_i)\right].\label{onn}\ee
It is not difficult to understand why these sums factorize.
The only difference between the $n!$ Feynman diagrams is  the order
of emission of the bosons. When summed over all possible orderings,
each boson is allowed to be emitted independently
at any place along the source,
hence factorization results. Eikonal approximation is needed to
derive factorization because we have implicitly assumed 
the source not to recoil in the above
argument. Otherwise it does make a difference whether a boson
is emitted before or after a recoil takes place.

\section{Nonabelian Bosons}
The algebra becomes more complicated in the nonabelian case because
the vertex factors $\s_i$ do not commute with one another. The
case for $n=2$ is still trivial to work out, but the combinatorials
for an arbitrary $n$ are fairly involved, so we will be content
just to state the final result here. 

\subsection{$n=2$}
By adding and subtracting $a^*[21]\s_1\s_2$ or $a[21]\s_1\s_2$, we get
\be
M^*_2&=&a^*[12]\s_1\s_2+a^*[21]\s_2\s_1\nonumber\\
&=&{1\over\o_1+i\e}{1\over\o_1+\o_2+i\e}
\s_1\s_2+{1\over\o_2+i\e}{1\over\o_1+\o_2+i\e}\s_2\s_1\nonumber\\
&=&{1\over\o_1+i\e}{1\over\o_2+i\e}\s_1\s_2
+{1\over\o_2+i\e}{1\over\o_1+\o_2+i\e}[\s_2,\s_1],\label{naoff2}\\
M_2&=&a[12]\s_1\s_2+a[21]\s_2\s_1\nonumber\\
&=&-2\pi i\delta(\o_1+\o_2)\left\{
{1\over\o_1+i\e}\s_1\s_2+{1\over\o_2+i\e}\s_2\s_1\right\}\nonumber\\
&=&(-2\pi i)^2\delta(\o_1)\delta(\o_2)\s_1\s_2-2\pi i\delta(\o_1+\o_2)
{1\over\o_2+i\e}[\s_2,\s_1].
\label{naon2}
\ee
\subsection{Atoms}
An {\it atom} $A_c^*$ or $A_c$
 is a Feynman amplitude whose product of vertices
has been replaced by their multiple commutators:
\be
A_c^*[s_1s_2\cdots s_n]&=&a^*[s_1s_2\cdots s_n][\s_{s_1},
[\s_{s_2},[\cdots[\s_{s_{n-1}},\s_{s_n}]\cdots]]],\label{atomoff}\\
A_c[s_1s_2\cdots s_n]&=&a[s_1s_2\cdots s_n][\s_{s_1},
[\s_{s_2},[\cdots[\s_{s_{n-1}},\s_{s_n}]\cdots]]].\label{atomon}\\
\ee
In terms of these, eqs.~(\ref{naoff2}) and (\ref{naon2})
can be written 
\be
M_2^*&=&A_c^*[1]A_c^*[2]+A_c^*[21],\nonumber\\
M_2&=&A_c[1]A_c[2]+A_c[21].\label{atom2}
\ee
It will be convenient to write a single notation for the products
of atoms by merging the symbols together. For example, we shall write
$A_c[1|2]=A_c[1]A_c[2]$, and $A_c[51|342|6]=A_c[51]A_c[342]A_c[6]$,
with the vertical bar $|$ indicating where products should
occur. We shall call the vertical bar a {\it cut} and
these single $A_c$'s as {\it cut amplitudes} \cite{FHL}.
In terms of this notation we can rewrite eq.~(\ref{atom2})
in terms of the cut amplitudes as follows:
as
\be
M_2^*&=&A_c^*[1|2]+A_c^*[21],\nonumber\\
M_2&=&A_c[1|2]+A_c[21].\label{atom2a}
\ee

\subsection{Decomposition theorem for arbitrary $n$}
The atomic decomposition for a general BE-symmetrized amplitude
$M_n^*$ or $M_n$ is 
\be
M^*_n&=&\sum_{\{s\}\in S_n}^{n!}A_c^*[s_c],\nonumber\\
M_n&=&\sum_{\{s\}\in S_n}^{n!}A_c[s_c].\label{sumatom}
\ee
In other words, we simply replace the Feynman amplitudes $A^*[s]$ and
$A[s]$ in (\ref{sum})
by the corresponding {\it cut amplitudes} $A^*_c[s_c]$ and $A_c[s_c]$,
where $[s_c]$ is $[s]$ with some vertical bars inserted.

We must still specify how to put the vertical cuts inside the symbol 
$s=s_1s_2\cdots s_n$ to get $s_c$. 
The rule is simple though the proof is not 
\cite{LL1}. Starting from the left, a vertical cut is put after
$s_i$ iff $s_i<s_j$ for all $j>i$. It is trivial to see that
eq.~(\ref{atom2a}) satisfies this rule. For further illustration,
the atomic decomposition for $n=3$ is
\be
M_3&=&A_c[1|2|3]+A_c[1|32]+A_c[21|3]+A_c[231]+A_c[31|2]+A_c[321].
\label{atom3a}
\ee
\subsection{Abelian decomposition}
For abelian vertices $\s_i$ all commutators vanish, so only one-boson
atoms survives. The decomposition (\ref{sumatom}) has only one term,
$A^*_c[1|2|3|\cdots|n]$ or $A_c[1|2|3|\cdots|n]$. In other words,
it factorizes in the way given by eqs.~(\ref{offn}) and (\ref{onn}).

For offshell amplitudes, this complete 
factorization leads to a Poissonian multiplicity 
distribution for the production of photons. It enables the amplitudes
of all orders to be summed up to an exponential form, thus allowing an
eikonal and a geometrical interpretation \cite{CW,EIK,GLAU}. 
It is also this exponential
form that allows the infrared divergences to be cancelled \cite{IR}.

For onshell amplitudes, the appearance of $\delta(\o_i)=\delta(2p\.k_i)$
indicates for example
that the photons bremstrahlung from electron-electron scattering
must be emitted in the forward-backward directions. 

Since the abelian cases are rather well known, we shall now pass on
to the nonabelian situation.

\subsection{Spacetime and colour interferences}
\subsubsection{spacetime interference}
Each onshell amplitude consists of a $\delta$-function 
(see eq.~(\ref{onshell})), whose argument
is the sum of $\o$'s of the bosons. 
When a BE-symmetrized {\it onshell} amplitude $M_n$ is decomposed into the
sum of Feynman amplitudes, as in (\ref{sum}), the dependence on the
remaining $(n-1)$ $\o$'s is of the form $\prod(\sum\o_i)^{-1}$
(see (\ref{onshell})), which has a broad distribution in the $\o$
variables. In contrast, when $M_n$ is decomposed in 
terms of
sums of products of {\it atoms}, as in (\ref{sumatom}), $(\alpha-1)$
additional $\delta$-functions appear in the term with $\alpha$ atoms. 
Just as in Sec.~3.1, these 
$\delta$-functions may be thought of as peaked interference
patterns produced by the coherent addition of the various broad Feynman
amplitudes. 

Constructive interference occurs when the argument of the
 $\delta$-function is zero, and complete destructive interference 
occurs whenever the argument is nonzero. In a high-energy scattering
amplitude, is {\it the overall interference effect} constructive, or
 destructive? At least in the two examples being discussed in the
Introduction, 
both of them appear to be destructive for the following reasons.

Schematically the $\ln(s)$ factors found in the elastic scattering
Feynman amplitudes come from integrations of the form
$\int^sd\o/\o=\ln(s)$. With interference, some $1/\o$'s are
changed to $\delta(\o)$'s. Since
$\int d\o\delta(\o)=1$, the $\ln(s)$ dependence disappears. 
This shows that cancellation of $\ln(s)$ factors in elastic scattering
is a BE {\it destructive interference} effect.

For QED (\ref{sumatom}) has $(n-1)$ atoms and $(n-1)$ additional
$\delta$-functions. This is why a complete cancellation of $\ln(s)$
occurs in electron-electron scattering mediated by the multiple exchange of
photons. For QCD atoms of all sizes appear, leaving
behind fewer $\delta$-functions and less $\ln(s)$ cancellations.
In particular, there are no extra $\delta$-functions in the term
with a single atom, so that term gives rise to
 no $\ln(s)$ cancellations. Eventually these uncancelled powers of
$\ln(s)$ will sum up to be a power of $s$, bringing reggeization
to the gluon. Since
atoms always carry octet colour, this happens only in the octet
channel, which is why the gluon may reggeize but the photon may not.

Other terms contain more than one atom so additional $\delta$-functions
are present and some amount of $\ln(s)$ cancellations occur 
\cite{FHL,FL1}.

\subsubsection{Colour interferrence}
The multiple commutators appearing in an atom may be interpreted as
exhibiting  colour interference. To see how this comes about let
us compare this with a Feynman amplitude, where the `colour matrices'
$\s_i$ appear as ordinary products. If each matrix is coupled to a
boson in the `adjoint representation', then a Feynman amplitude with
$n$ vertices is coupled to $n$ bosons whose total colour spreads
over a wide range, a range that involves
 any colour obtainable by coupling $n$
adjoint colours. In contrast, when these $n$
colour matrices appear as a multiple commutator, only adjoint 
representation survives. The adjoint commutator may therefore
be thought of as a {\it colour interference} pattern,
obtained by coherently adding up several broad colour distributions
to yield a pattern which is
sharply peaked at the adjoint colour and zero at all other
colours.

It is this
colour interference that averts a potential
disaster \cite{LL2} in the inelastic
pi-nucleon process $\pi+N\to (n-1)\pi+N$ in large-$N_c$ QCD.

Imagine using Fig.~1 to describe this process. It is known that
this Feynman diagram is of order $N_c^{n/2}$. On the other hand,
it is also known that the full amplitude, obtained by summing over
the $n!$ permutations of the pions, must behave like 
$N_c^{1-n/2}$ \cite{NC}.
What is the mechanism for $n-1$ powers of $N_c$ 
to be cancelled out in the sum?
The answer turns out to be `colour interference', at least for tree
diagrams \cite{LL2}. A similar cancellation must take
place for loop diagrams as well, 
but I have not yet been able to figure out
how it works there.

A colourless nucleon is made up of $N_c$ quarks. Even if all these quarks
are in the $S$ state, arbitrary spin and isospin alignments for the
$N_c$ quarks are allowed, thus producing many nucleon resonances
with spins and isospins ranging from $\h$ to $\h N_c$.
It turns out to be these high-spin/isospin  resonances that give
rise to the unacceptably
large contribution $N_c^{n/2}$ in the amplitude. If somehow these
contributions can be suppressed in the sum, then there is a  chance to
obtain the right answer $N_c^{1-n/2}$.

In the rest system of the heavy nucleon,
whose mass $M$ is proportional to $N_c$, $\o_i=2p\.k_i$
is $2M$ times the pion energy in this frame. Our notation is
such that the $\o$ for the
outgoing pions are positive and the $\o$ for the single incoming pion
is negative. Since the pions are massive, it is impossible for
the sum of any number of {\it outgoing} $\o_i$'s to vanish,
so any term that involves a $\delta$-function with this sum as
its argument would disappear. As a result,
only single-atom terms
survive in the atomic decomposition (\ref{sumatom}). However,
the `colour' (actually spin and isospin here) of a single atom
is composed of multiple commutators, all of them having only
the adjoint `colour' (meaning singlet and triplet spins and isospins here).
This makes it impossible for nucleon resonances of high spin/isospin to
contribute, and it can be shown that this makes the total sum to 
behave correctly like $N_c^{1-n/2}$ \cite{LL2}.

\subsection{Complementarity between spacetime and colour interferences}
The sharpness of spacetime interference can be measured by the number
of additional $\delta$-functions present. The sharpness of colour 
interference can be measured by the number of commutators around.
Since in every term of the atomic decomposition (\ref{sumatom}),
the sum of these two is always equal to $n-1$,  a complentarity
exists between the two kinds of interferences. If one is large,
then  the other is small, and vice versa.

\subsection{Advantage of atomic decomposition and nonabelian cut diagrams}
The decomposition formula (\ref{sumatom}) can be considered as a
resummation formula for  eq.~(\ref{sum}),
with both
spacetime and colour interferences, and hence cancellations, automatically
built in. 

Most multliloop Feynamn
diagrams cannot be computed analytically, though in the 
presence of a large parameter ({\it e.g.,} $\ln(s)$ or $N_c$)
one might hope in certain cases to compute
it approximately. Even so one can expect to compute it only up to
the leading approximation, {\it viz.,} the leading power of $\ln(s)$
or $N_c$. If such leading contributions to individual Feynman diagrams
cancel out in the sum, we would be left with no means to calculate
other than tackling the subleading contributions, which is very difficult.
That is not all, for the subleading
and the sub-subleading contributions may also be cancelled.
Such is the case in the elastic scattering of two electrons 
where all powers of $\ln(s)$ are cancelled,
and is also the case for the $\pi N$ inelastic
scattering problem where $(n-1)$ powers of $N_c$ are cancelled. 

The atomic decomposition now offers a new way to do such calculations.
Since all interference and cancellation effects are already built into
the atoms, or the factorization, no further unwanted cancellations will
occur if we take the large $\ln(s)$ or large $N_c$ limit. 
This atomic decompostion can actually be implemented in a graphical way,
by making trivial changes to the Feynman diagrams to turn them into
{\it nonabelian cut diagrams} \cite{FHL,FL1} suitable for such calculations.
As a matter of fact, besides not having to deal with cancellations at
the end, individual nonabelian
cut diagrams are actually easier to calculate than individual
Feynman diagrams.

\section{Conclusion}
It is somewhat surprising that many seemingly unrelated phenomena can
be understood by the {\it factorization} and {\it interference} effects
inherent in BE symmetrization. It is interesting that by taking this
symmetrization early in the calculation one can produce a new and
powerful calculational scheme in terms of
the {\it nonabelian cut diagrams}. The only approximation needed
to obtain these generic and common results is to have sources to be
energetic or massive. Many physical phenomena seem to fall into this
category, so we are hopeful that the technique and the understanding
developed here can be used to study a variety of unsolved nonabelian problems.

\end{document}